\documentclass[12pt,dvips]{article}

\makeatletter
\renewcommand{\section}{\@startsection{section}{1}{0in}
	{0.4\baselineskip}{0.1\baselineskip}{\Large\bf}}
\renewcommand{\subsection}{\@startsection{subsection}{2}{0in}
	{0.25\baselineskip}{-\baselineskip}{\large\bf}}
\renewcommand{\subsubsection}{\@startsection{subsubsection}{3}{0in}
	{0.1\baselineskip}{-\baselineskip}{\normalsize\bf}}
\makeatother
\begin{document}

%
\makeatletter\newcommand{\ps@icrc}{
\renewcommand{\@oddhead}{\slshape{ICRC 2001 Posters
(updated comment)}\hfil}}
\makeatother\thispagestyle{icrc}
%
%

\begin{center}
%
~ 

{\LARGE \bf Deformed Lorentz Symmetry\\ 
and High-Energy Astrophysics (II)}
\end{center}

\begin{center}
%
%
~ 

{\bf Luis Gonzalez-Mestres$^{1,2}$}\\

~ 

{\it $^{1}$ L.A.P.P., CNRS-IN2P3, B.P. 110, 74941 Annecy-le-Vieux Cedex, France \\
$^{2}$ L.I.G.A., CNRS-ENS, 45 rue d'Ulm, 75230 Paris Cedex 05, France }
\end{center}

~ 

\begin{center}
{\large \bf Abstract\\}
\end{center}
\vspace{-0.5ex}
%
%
Lorentz symmetry violation (LSV) 
can be generated at the Planck scale, or at some
other fundamental length scale, and
naturally preserve Lorentz symmetry
as a low-energy limit (deformed Lorentz symmetry, DLS). 
DLS can
have important implications for ultra-high energy cosmic-ray physics
(see papers physics/0003080 - hereafter referred to as I - , astro-ph/0011181 and astro-ph/0011182, and references quoted in these papers). A crucial
question is how DLS can be extended to a deformed Poincar\'e symmetry (DPS), and what can be the dynamical origin of this phenomenon. We
discuss recent proposals to identify DPS with a symmetry incorporating the Planck scale (doubly special relativity, DSR) and suggest new ways in this direction. Implications for models of quadratically deformed relativistic kinematics (QDRK) and linearly deformed relativistic kinematics (LDRK) are also discussed.

~ 

This paper updates and further develops some aspects
of contributions to the
ICRC 2001 Conference, Hamburg August 2001 (Posters: 0G 092 , OG 093 and HE 313).
~
~
\vspace{1ex}

~

%
%
~ 

\section{Deformations and non-unitary operators}
\label{relativity.sec}

~
~ 

Are DLS and DPS broken symmetries or real symmetries? The algebraic approach tends to consider them as real symmetries incorporating the Planck scale as an extra fundamental parameter (see f.i. Lukierski and Nowicki, 2002, and references therein; Amelino-Camelia, Benedetti and D'Andrea, 2002; Amelino-Camelia, 2001 and 2002; Bruno, Amelino-Camelia and Kowalski-Gilman, 2001). This is in itself not required from a phenomenological point of view, but must be discussed and explored.

~ 

When completing this work, we recently found a paper (Magueijo and Smolin, 2002) where some basic ideas and models present similarities with those developed here. We refer to it and to our previous papers (Gonzalez-Mestres, 2000 a, b and c) for a description of the formalism used. It must be emphasized, however, that our physical assumptions are different from those of Magueijo and Smolin. A crucial question for the consistency of any deformed relativistic kinematics (DRK) is how the deformation depends on mass for large bodies.

~ 

The scenario discussed in the present paper is as follows. We assume that a reference dynamical system exists with a standard Poincar\'e group. But the real physical reference system is a different one, where the energy and momentum conserved in the physical space-time do not correspond to those of the standard underlying Poincar\'e group. Contrary to Magueijo and Smolin, we assume that in the physical reference system energy and momentum conservation is preserved and that energy is additive, as well as momentum, for a system of several non-interacting bodies. We also tacitly assume the existence of a physical vacuum rest frame, where everything is formulated in spite of the formal underlying Lorentz invariance. Following Poincar\'e (1895, 1901, 1905), we interpret special relativity as the impossibility to identify the vacuum rest frame by available physical measurements. But cosmology clearly seems to suggest the existence of a preferred reference frame (see, f.i., Peebles, 1993), and this circonstance must be fully understood from the point of view of Lorentz symmetry.

~ 

Following notations similar to (Magueijo and Smolin, 2002), we assume that the transition between the (possibly fundamental) symmetric reference system and the (experimentally natural) physical one is driven by a singular non-unitary operator:

\equation
U ~ (p_0 ~ , ~ p_i ~, ~ T ~ , ~ T_0 ~ , ~ a) 
\endequation
\noindent
transforming a physical state $|\psi> ~ (physical)$ into an object in the symmetric reference system, $|\psi> ~ (symmetric)$ :

\equation
|\psi> ~ (symmetric) ~ = ~ U ~ |\psi> ~ (physical)
\endequation
\noindent
and, forgetting about speed of light and Planck constant, $p_0$ is the energy, $p_i$ stands for momentum in the three space directions, $T$ is an intrinsic time scale associated to each particle, $T_0$ is a universal time scale and $a$ is the fundamental length scale (e.g. the Planck length). We ignore the problems raised by singularities but, as it will be discussed later, it is likely that in most cases an equivalent unitary operator can be found. We can then write for the energy-momentum quadrivector operator:

\equation
p_{\alpha} ~ (symmetric) ~ = ~ U ~ p_{\alpha} ~ (physical) ~ U^{-1}
\endequation
\noindent
where "physical" stands for the measured energy and momentum, and "symmetric" for energy and momentum in the ideal symmetric frame. The index $\alpha ~ = ~ 0 ~ , ~ 1 ~ , ~ 2, ~ 3$ corresponds to energy and the three momentum coordinates with the metrics (+, -, -, -). 
\vskip 2mm
In the symmetric system, the equation of motion for free particles:

\equation
p_{0}^2 ~ (symmetric) ~ = ~ \Sigma _{i=1} ^3 ~ p_{i}^2 ~ (symmetric) ~ + ~ m^2
\endequation
\noindent
$m ~ = ~ T^{-1}$ ~ being the mass. 

~ 

We shall be interested in models where, for instance:

\equation
U (p_0 ~ , ~ p_i ~, ~ T ~ , ~ T_0 ~ , ~ a) ~ = exp ~ [-1/2 ~ ~ ln ~ (1 ~ - ~ \epsilon ~ p_0^2 ) ~ ~D_{space}]
\endequation
\noindent
$\epsilon $ being a parameter that will be determined below and $D_{space} ~ = p^i ~ \partial /\partial p^i$ (the sum over space indices is understood) the space dilatation generator. The symbol $ln$ stands for neperian logarithm. $U$ is obviously a singular operator. We then get:

\equation
p_{0} ~ (symmetric) ~ = ~ p_{0} ~ (physical)
\endequation
\equation
p_{i} ~ (symmetric) ~ = ~ p_{i} ~ (physical) 
~ [1 ~ + ~ \epsilon ~ p_0^2 ~ (physical)]^{-1/2}
\endequation

We take in what follows $p_{\alpha}$ for $p_{\alpha} ~ (physical)$ and use the equation of motion in the symmetric reference system expressed, in the physical system, by the equation:

\equation
p_{0}^2 ~ = ~ p^2 ~ [1 ~ + ~ \epsilon ~ p_0^2]^{-1} ~ + ~ m^2
\endequation
\noindent
equivalent to: 

\equation
p_{0}^2 ~ - ~ m^2 ~ = ~ p^2 ~ [1 ~ - ~ f ~ (p^2)]^{-1}
\endequation
\noindent
where $p^2 ~ = ~ \Sigma _{i=1} ^3 ~ p_i^2$ and:

\equation
2 ~ f ~ (p^2) ~ ~ = ~  ~ 1 ~ - ~ \epsilon ~ m^2 ~ - ~ [(1~ - ~ \epsilon ~ m^2)^2 ~ + ~ 4 ~ \epsilon ~ (p^2 ~ + ~ m^2)]^{1/2}
\endequation

This expression defines the hamiltonian $H$ in the physical reference system. For $\epsilon ~ p^2 ~ \ll ~ 1$, it gives:

\equation
p_{0}^2 ~ - ~ m^2 ~ \simeq ~ p^2 ~ (1 ~ - ~ \epsilon ~ m^2 ) ~ - ~ \epsilon p^4 
\endequation
\noindent
which is similar to the QDRK (quadratically deformed relativistic kinematics) considered in I. Another possibility would be to take:

\equation
ln ~U~ (p_0 ~ , ~ p_i ~, ~ L ~ , ~ L_0 ~ , ~ a)~~ =~~ -1/2 ~ ~ ln ~ [1 - ~ \epsilon (p^2 ~ - ~ m^2)] ~ D_{time}
\endequation
\noindent
where $D_{time} ~ = ~ p_0 ~ \partial /\partial p_0$ is the time dilatation generator, and $L~=~c~T$ and $L_0~=~c~T_0$ are constants. We then get:

\equation
p_{0}^2 ~ = ~ p^2 ~ + ~ m^2 (1 ~ + ~ \epsilon ~ m^2) ~ - ~ \epsilon ~ p^4 
\endequation
\noindent
which coincides exactly (up to a trivial redefinition of mass) with the main QDRK model discussed in I. The differences between (11) and (13) are physically unobservable in practical cases. For a proton, assuming that the deformation term becomes of the same order as the mass term at $E ~ \approx ~ 10^{19} ~ GeV$, $\epsilon ~ m^2$ would have to be $\approx ~ 10^{-40}$ , so that the change in critical speed introduced by the term proportional to $\epsilon ~ m^2 ~ p^2 $ in (11) cannot produce any effect measurable by existing techniques (it is much smaller than those considered by Coleman and Glashow, 1997). A similar situation was already found when studying the sine-Gordon equation discretized with a one-dimensional space lattice with spacing $a$ (Gonzalez-Mestres, 1997 a and b). A perturbative expansion in powers of  $p ~ a$  for the lattice equation brought the same kind of deformation of the critical speeds.

~ 

It is also possible to obtain useful versions of LDRK with simple choices, such as:

\equation
ln ~U~ (p_0 ~ , ~ p_i ~, ~ T ~ , ~ T_0 ~ , ~ a)~ =~ -1/2 ~ ~ ln ~ [1 - ~ \mu (p^2 ~ + ~ m^2)^{1/2}] ~ D_{time}
\endequation
\noindent
leading to:
\equation
p_{0}^2 ~ = ~p^2 ~ + ~ m^2~ -~\mu~ ( ~p^2 ~ + ~ m^2)^{3/2}
\endequation
\noindent
where $\mu $ is the coefficient of the deformation term.

~ 

For fermions obeying a Dirac equation:

\equation
[\gamma _{\alpha } ~ p^{\alpha } ~ (symmetric) ~ - ~ m]~ |\psi> ~ = ~ 0
\endequation
\noindent
where the $\gamma $'s are Dirac matrices,
we can use $U$ such that:

\equation
ln ~U~ (p_0 ~ , ~ p_i ~, ~ L ~ , ~ L_0 ~ , ~ a)~~ =~~ - ~ ~ ln ~ [1 - ~ \epsilon ~ (p^2 ~ - ~ m ~ \gamma _i ~ p^i ~ + ~m^2)] ~ ~ D_{time}
\endequation
\noindent
leading to the fermionic QDRK:
\equation
[\gamma _{\alpha } ~ p^{\alpha } ~ + ~ \epsilon ~ p^2 ~ \gamma _i ~ p^i ~ - ~ m ~ (1~ - ~ \epsilon ~ m^2)]~ |\psi> ~ = ~ 0
\endequation
\noindent

A LDRK for fermions is obtained using $U$ such that:

\equation
ln ~U~ (p_0 ~ , ~ p_i ~, ~ L ~ , ~ L_0 ~ , ~ a)~~ =~~ - ~ ~ ln ~ [1 - ~ \mu ~ (\gamma _i ~ p^i ~ - ~m)] ~ ~ D_{time}
\endequation
\noindent
leading to:

\equation
[\gamma _{\alpha } ~ p^{\alpha } ~ + ~ \mu ~ p^2 ~ - ~ m ~ (1~ + ~ \mu ~ m^2)]~ |\psi> ~ = ~ 0
\endequation
\noindent

It must be noticed that the relation between space and time, such as they are defined in the physical reference system, and their equivalents in the symmetric reference system, i.e:

\equation
[t~, ~ x_i] ~ (symmetric) ~ ~ = ~ ~ U ~ [t ~ , ~ x_i] ~ (physical) ~ U^{-1}
\endequation
\noindent
is highly nontrivial. In particular, the two sets of variables do not commute with each other (but each set of space-time coordinates does). In each reference system, energy-momentum conservation holds with respect to the space-time of the system under consideration.

~
~ 

\section{Additivity}
\label{additivity.sec}
\vspace{0.5ex}

~

As previously stated, we assume that in the physical reference system energy and momentum are additive for a system of several non-interacting bodies. We will follow here and approach similar to (I). For a set of two large non-interacting objects of masses $m_1$ and $m_2$ travelling at the same speed, the law $\epsilon \propto m^{-2}$, assigning a mass $M ~ = ~ m_1 ~ + ~ m_2$ to the two-particle system, guarantees the consistency of the deformed kinematics (Gonzalez-Mestres, 1997 a and b and 1998). Writing, for large bodies:

\equation
\epsilon ~ \simeq ~ \lambda ~ T^2 ~ T_0^{-2} a^2 ~ = ~ \epsilon _0 ~ m^{-2}
\endequation
\noindent
where $\lambda $ and $\epsilon _0 $ are constants, equation (13) becomes : 

\equation
p_{0}^2 ~ = ~ p^2 ~ + ~m^2 ~ (1 ~ + ~ \epsilon _0 ) - ~ \epsilon _0 ~ m^{-2} ~ p^4
\endequation
\noindent
and:

\equation
p_0 ~ m^{-1} ~ = ~ g ~ (p^2 ~ m^{-2})
\endequation
\noindent
where:

\equation
g (z) ~ = ~ (1 ~ + ~ \epsilon _0 ~ + ~ z ~ + ~ \epsilon _0 ~ z^2)^{1/2}
\endequation
\noindent
so that the velocity vector $v_i ~ = ~ \partial H/\partial p^i $ is a function of $p_i ~ m^{-1}$ and $p^2 ~ m^{-2}$. The situation is the same for the deformed Dirac equation in the case of fermions. This confirms that consistency for systems of large bodies is reached with the following assumptions:

~ 

- The total mass of a system of N non-interacting bodies travelling together at the same speed is given by the sum of the masses of these objects.

~ 

- In the physical system, the total energy and momentum of a N-body non interacting system are given the sum of their individual energies and momenta.

~ 

- For large bodies, the deformation parameter $\epsilon $ follows a universal law: $\epsilon ~ \simeq ~ \epsilon _0 ~ m^{-2}$ where $ \epsilon _0 $ is a universal parameter.

~

The value of the constant $T_0$ is related to the transition from the elementary particle domain, where we can expect $\epsilon $ to depend weakly on the mass of the particle, to that of composite objects (f.i. nuclei and heavier bodies) where the asymptotic regime defined by (22) should be reached. We therefore expect $T_0$ to be in the range 0.1 to 1 $GeV$.

~ 

Similarly, for LDRK, we replace our third assumption by the statement:

~ 

- For large bodies, the deformation parameter $\mu $ follows a universal law: $\mu ~ \simeq ~ \mu _0 ~ m^{-1}$ where $ \mu _0 $ is a universal parameter given by the expression $\mu _0 ~ = ~ \rho ~ T_0^{-1} ~ a$ and $\rho $ is a constant.

~
~

\section{On the use of the new operators}
\label{renor.sec}
\vspace{0.5ex}

~

An interesting question is how far can we use this kind of singular operators, and how many physical problems they can help to solve. Can they be a way to renormalize field theories or make them finite? We start here the discussion by showing how it is formally possible to transform a world with chiral symmetry and only massless particles into a world where particles have masses, and how the operator $U$ can be made unitary. 

~ 

If all particles are massless, we have a universal equation:

\equation
p_0^2 ~ (symmetric) ~ - ~ p^2 ~ (symmetric) ~ = ~ 0
\endequation
\noindent
but, if the singular operator $U$ transforms $p_{\alpha }$ into: 

\equation
p_{\alpha } ~ (symmetric) ~ = ~ p_{\alpha } ~ (p_0^2 ~ - ~ p^2 ~ -~ m^2)^{1/2} ~ (p_0^2 ~ - ~ p^2)^{-1/2}
\endequation
\noindent
then, equation (26) turns into:

\equation
p_0^2 ~ (physical) ~ - ~ p^2 ~ (physical) ~ - m^2 ~ = ~ 0
\endequation
\noindent

To obtain this result, we can use an operator $U$ defined by:

\equation
ln ~ U ~ ~ = ~ ~ 1/2 ~ ~ ln ~ [1 ~ - ~ m^2 ~ (p_0^2 ~ - ~ p^2 ~ + ~ \eta _1 ~ m^2)^{-1} ~ + ~ \eta _2] ~ ~ D
\endequation
\noindent
where $D ~ = p_ {\alpha } ~ \partial/\partial p_ {\alpha }$ (the sum over the index ${\alpha }~ =~ 0~ , ~ 1~ , ~ 2~ , ~ 3)$ is understood) is the dilatation operator and $\eta _1$ , $\eta _2$ are very small positive numbers introduced to avoid on-shell singularities in intermediate steps. Their systematic use is understood in the rest of the chapter, the limit $\eta _1$ , $\eta _2$ $\rightarrow ~ 0$ being equally understood at the end of the calculation. We consider $m$ as a Lorentz-invariant internal energy-momentum scale of each particle so that $[D ~ , ~ m] ~ = ~ m$ , or add a fifth space-time dimension (including it in the definition of $D$) to produce the same effect. Then, the expression $ ln ~ [1 ~ - ~ m^2 ~ (p_0^2 ~ - ~ p^2)^{-1}]$ is Lorentz and dilatation invariant and the result (27) is obtained. A similar transformation can be used for fermions, with:

\equation
ln ~ U ~ ~ = ~ ~ 1/2 ~ ~ ln ~ [1 ~ - ~ m ~ (\gamma _{\alpha } ~ p^{\alpha })^{-1}] ~ ~ D
\endequation
\noindent
where the same convention for $m$ is taken as before, in order to turn the equation:

\equation
\gamma _{\alpha } ~ p^{\alpha }~ |\psi> ~ = ~ 0
\endequation
\noindent
into: 

\equation
(\gamma _{\alpha } ~ p^{\alpha } ~ - ~ m)~ |\psi> ~ = ~ 0
\endequation
\noindent

Replacing energy and momentum by their gauge covariant equivalents, and assuming the gauge invariance of $m$, nothing seems to prevent the use of similar techniques in the presence of gauge couplings. More generally, the operator $U$ can in principle simultaneously incorporate phenomena such as conventional symmetry breaking conserving standard Lorentz invariance and more unconventional physics accounting for the deformation of relativistic kinematics. Indeed, writing for exemple:

\equation
ln ~ U ~ ~ = ~ ~ 1/2 ~ ~ ln ~ [1 ~ - ~ (m^2 ~ - ~ \epsilon ~ p^4) ~ (p_0^2 ~ - ~ p^2)^{-1}] ~ ~ D
\endequation
\noindent
and assuming, as before, that $m$ transforms like energy or momentum under dilatations, and similarly for $\epsilon ^{-1/2}$, we recover a simple QDRK equation:

\equation
p_{0}^2 ~ = ~ p^2 ~ + ~ m^2 ~ - ~ \epsilon ~ p^4 
\endequation
\noindent

To obtain this result, we can introduce extra momentum dimensions for $m$ and $\epsilon ^{-1/2}$, and use a generalized dilatation operator in the resulting six-dimensional space-time. The interpretation of $\epsilon ^{-1/2}$ as a momentum may be linked to the breaking of a six-dimensional space-time symmetry. Similar techniques can be used for LDRK and fermions.

~

At this stage, it seems necessary to discuss the question of the unitarity of $U$. Taking, for instance, an expression like (33), we can build an equivalent unitary operator $U'$ by writing:

\equation
4 ~ ln ~ U' ~ ~ = ~ ~ ln ~ [1 ~ - ~ (m^2 ~ - ~ \epsilon ~ p^4) ~ (p_0^2 ~ - ~ p^2)^{-1}] ~ ~ D ~ ~ + ~ ~ \partial/\partial p_ {\alpha } ~ p_ {\alpha } ~ ~ ln ~ [1 ~ - ~ (m^2 ~ - ~ \epsilon ~ p^4) ~ (p_0^2 ~ - ~ p^2)^{-1}]
\endequation
\noindent
which, using the scale invariance of the expression $1 ~ - ~ (m^2 ~ - ~ \epsilon ~ p^4) ~ (p_0^2 ~ - ~ p^2)^{-1}$ with our generalization of $D$, is equivalent to: 

\equation
ln ~ U' ~ ~ = ~ ~ 1/2 ~ ~ ln ~ [1 ~ - ~ (m^2 ~ - ~ \epsilon ~ p^4) ~ (p_0^2 ~ - ~ p^2)^{-1}] ~ ~(D ~ - ~ d/2)
\endequation
\noindent
where $d$ is the number of space-time dimensions under consideration (4 or 6) and $U$ is unitary provided $1 ~ - ~ (m^2 ~ - ~ \epsilon ~ p^4) ~ (p_0^2 ~ - ~ p^2)^{-1}$ is made real and positive by the regulators $\eta _1$ , $\eta _2$ as before. We then get:

\equation
U' ~ ~ = ~ ~ [1 ~ - ~ (m^2 ~ - ~ \epsilon ~ p^4) ~ (p_0^2 ~ - ~ p^2)^{-1}]^{-d/4} ~ ~ U
\endequation
\noindent 
and the extra factor has no practical effect on the energy-momentum operators, as it commutes with $D$ and with any fonction of energy and momentum. Similar techniques apply to all versions of $U$ considered in the present paper, provided we assume that $m$ , $\epsilon ^{-1/2} $ and $\mu ^{-1}$ behave like momenta under dilatation or the four-dimensional space-time. $D$ can extended to the above mentioned six-dimensional space in order to produce the same result. Actually, in all relevant situations we can always define dilatations in such a way that the expression multiplying $D$ in the exponential form of $U$ commutes with $D$, so that an equivalent unitary operator $U'$ can be built and the "physical" states can be related to the "symmetric" ones by the expression: 

\equation
|\psi '> ~ (symmetric) ~ ~ = ~ ~ U' ~ |\psi> ~ (physical)
\endequation
\noindent 
and more generally, to build the $U'$ operator, if (taking QDRK):

\equation
ln ~ U ~ = ~ ~ h ~ (p_0 ~ p_i ~ , ~ m ~ , ~ \epsilon ^{-1/2}) ~ ~ D
\endequation
\noindent 
where $h$ is a real fonction of the six momenta, $D$ is defined in the six-dimensional space and $h$ is an invariant under the six-dimensional $D$ , we just define:

\equation
2 ~ ln ~ U' ~ = ~ ~ ln ~ U ~ - ~ (ln ~ U)^{\dagger } ~ ~ = ~ ~ h ~ (p_0 ~ p_i ~ , ~ m ~ , ~ \epsilon ^{-1/2}) ~ ~ D ~ + ~ \partial /\partial p_{\beta} ~ p_{\beta} ~ h ~ (p_0 ~ p_i ~ , ~ m ~ , ~ \epsilon ^{-1/2})
\endequation
\noindent
where the index $\beta $ corresponds to the six space-time dimensions and sum over $\beta $ is understood. Using the commutation relation between $ p_{\beta}$ and $\partial /\partial p_{\beta}$, we get for QDRK:

\equation
ln ~ U' ~ = ~ ~ h ~ (p_0 ~ p_i ~ , ~ m ~ , ~ \epsilon ^{-1/2}) ~ ~ (D ~ - ~ d/2)
\endequation
\noindent 
where $d$ = 6 in this specific case, and:

\equation
U' ~ = ~ ~ exp ~ [-d/2 ~ ~ h ~ (p_0 ~ p_i ~ , ~ m ~ , ~ \epsilon ^{-1/2})] ~ ~ U
\endequation
\noindent 
and again, $d/2$ = 3 . Similar considerations apply to LDRK and to fermions.

~

We also notice that DRK introduces, besides the fundamental length scale, another energy scale. Taking, for a proton in QDRK, $\epsilon_0 ~ \simeq ~ 0.1 ~ a^2$ and $a$ to be the Planck length, it is found (see I) that the deformation term equals the mass term at a transition energy scale around $E_{trans} ~ \approx ~ 10^{19} ~ GeV$. This transition is really dynamical and not just kinematical, as above $E_{trans}$ standard formulae for Lorentz contraction and time dilation do no longer hold (Gonzalez-Mestres, 1997 a and b). The standard parton picture also fails, but we can replace it by a new one if we admit that, indeed, $\epsilon ^{-1/2} $ behaves like a momentum and can be shared by the partons. Defining the six-dimensional momentum $P ~ =~ (p_0 ~ , ~ p_i ~ , ~ m ~ , ~ \epsilon ^{-1/2}) $ and considering $m$ as the rest energy of the 3-parton system, a parton picture can still be formulated if $P$ can be shared by partons carrying any set fractions of it such that their sum is 1 . Then, the deformation term $ \epsilon ~ p^4$ can be interpreted as:

\equation
\epsilon ~ p^4 ~ ~ = ~ ~ (\epsilon ~ p^2) ~ p^2 ~ ~ = ~ ~\Omega ^2~ p^2
\endequation
\noindent 
where $\Omega ~ = ~ \epsilon ^{1/2} ~ p$ is the tangent of an angle between two components of the six-dimensional momentum, and similarly for $\epsilon ^{1/2} ~ m$. The component in the direction of $\epsilon ^{-1/2} $ would then be of an order of magnitude close to Planck momentum, therefore much larger than $p$ in the physical situations under consideration. $\Omega $ would then be a very small angle, increasing linearly with $p$ , and $E_{trans}$ the energy scale at which the role of the sixth dimension becomes apparent. We conjecture that the energy scale $E_{trans}$ can play a crucial role in renormalization, as it is possible to use it to generate dimensionless quantities of order $\approx ~ 1 $ , such as $m ~ E_{Planck} ~ E_{trans}^{-2}$ which may help to regularize infinities without producing large numbers of order $ E_{Planck} ~ m^{-1}$. This question will be discussed in a forthcoming paper.

~ 

It should also be pointed out that the transformation initially suggested in (Magueijo and Smolin, 2001) had rather trivial effects on energy and momentum, and did not lead to cubic or quartic terms when expressing $E$ as a fonction of $p$. With our convention for the definition of dilatations and using $\mu $ as the deformation parameter, the ansatz by Magueijo and Smolin would have given:

\equation
p_{0}^2 ~ = ~ p^2 ~ + ~ m^2 ~ exp ~ [- ~ \mu ~ p_0]
\endequation
\noindent
which is an acceptable LDRK, and replacing as before $D$ by $D ~ - ~ d/2$
would have made the transformation unitary.

~ 

\section{Conclusion and comments}
\label{concl.sec}
\vspace{0.5ex}

~

The $U$ operators proposed by Magueijo and Smolin are a useful tool to manipulate and understand deformed relativistic kinematics. We have discussed here some possible improvements of this approach. But further work is required on the physical interpretation of the formalism. The fact that, in a given reference frame, DRK can be made equivalent to a relativistic equation via the $U'$ operator does not by itself imply that the situation is the same in other frames. It may happen, for instance, that the $U'$ (or $U$) operator is not universal and varies when a boost is performed in the "symmetric" system. Then, it would still be possible to distinguish between the vacuum rest frame and other inertial frames. Physics is defined only by the full set of inertial frames with the corresponding transformation laws, plus the full lagrangian or Hamiltonian in each frame. The so-called DSR, "doubly special relativity" (Amelino-Camelia, Benedetti and D'Andrea, 2002; Amelino-Camelia, 2001 and 2002; Bruno, Amelino-Camelia and Kowalski-Gilman, 2001) is a particular case of DRK. However, from a phenomenological point of view it is not obvious whether the difference between exact realizations of DSR and DRK models violating it will ever be testable through feasible experiments. Our 1997 phenomenological models (see, f.i. Gonzalez-Mestres, 1997 a to 1997 e) just conjectured the existence of a vacuum reference frame with a DRK, and assumed that the laboratory moves slowly with respect to this frame. This is in principle enough for present-day experimental applications, but it would be beautiful if it were not.

~ 

%
%
%
\vspace{1ex}
\begin{center}
{\bf References}
\end{center}
%
\noindent
Amelino-Camelia, G., 2001, Phys.Lett. B510, 255. \\
Amelino-Camelia, G., 2002, Int. J. Mod. Phys. D11, 35.\\
Amelino-Camelia, G., Benedetti, D. and D'Andrea, F., 2002, paper hep-th/0201045.\\
Bruno, R., Amelino-Camelia, G. and Kowalski-Gilman, J., Phys. Lett. B522, 133.\\
Coleman, S. and Glashow, S.L., 1997, Phys.Lett. B405, 249 , paper hep-ph/9703240. \\
Gonzalez-Mestres, L., 1997a, paper nucl-th/9708028.\\
Gonzalez-Mestres, L., 1997b, Proc. of the International
Conference on Relativistic Physics and some of its Applications, Athens June 1997, paper physics/9709006. \\
Gonzalez-Mestres, L., 1997c, paper physics/9704017.\\
Gonzalez-Mestres, L., 1997d, Proc. 25th ICRC, Vol. 6, p. 113, physics/9705031.\\
Gonzalez-Mestres, L., 1997e, paper physics 9706032.\\
Gonzalez-Mestres, L., 1998, Proc. COSMO 97, Ambleside September 1997,
World Scient., p. 568, paper physics/9712056. \\
Gonzalez-Mestres, L. 2000a, paper physics/003080 , referred to as I.\\
Gonzalez-Mestres, L. 2000b, paper astro-ph/0011181, Proceedings of the International Symposium on High Energy Gamma-Ray Astronomy, Heidelberg, Germany, June 26-30, 2000. \\
Gonzalez-Mestres, L. 2000c, paper astro-ph/0011182, same Proceedings.\\
Lukierski and Nowicki, 2000, paper hep-th/0203065. \\
Magueijo and Smolin, 2001, paper hep-th/0112090. \\
Magueijo and Smolin, 2002, paper gr-qc/0207085. \\
Peebles, P.J.E., 1993, "Principles of Physical Cosmology", 
Princeton University Press.\\
Poincar\'e, H., 1895, "A propos de la th\'eorie de M. Larmor",
L'Eclairage \'electrique, Vol. 5,  5.\\
Poincar\'e, H., 1901, "Electricit\'e et Optique: La lumi\`ere
et les th\'eories \'electriques", Ed. Gauthier-Villars, Paris.\\
Poincar\'e, H., 1905, "Sur la dynamique de l'\'electron", C.
Rend. Acad. Sci. Vol. 140, p. 1504.\\
\end{document}